# RPYS i/o: A web-based tool for the historiography and visualization of citation classics, sleeping beauties, and research fronts

Jordan A. Comins[1] and Loet Leydesdorff[2,*]

**Abstract**

Reference Publication Year Spectroscopy (RPYS) and Multi-RPYS provide algorithmic approaches to reconstructing the intellectual histories of scientific fields. With this brief communication, we describe a technical advancement for developing research historiographies by introducing RPYS i/o, an online tool for performing standard RPYS and Multi-RPYS analyses interactively (at http://comins.leydesdorff.net/). The tool enables users to explore seminal works underlying a research field and to plot the influence of these seminal works over time. This suite of visualizations offers the potential to analyze and visualize the myriad of temporal dynamics of scientific influence, such as citation classics, sleeping beauties, and the dynamics of research fronts. We demonstrate the features of the tool by analyzing—as an example—the references in documents published in the journal *Philosophy of Science*.

[1] Center for Applied Information Science, Virginia Tech Applied Research Corporation, Arlington, VA, United States; jcomins@gmail.com

[2] *corresponding author*; Amsterdam School of Communication Research (ASCoR), University of Amsterdam, PO Box 15793, 1001 NG Amsterdam, The Netherlands; loet@leydesdorff.net

# 1. Introduction

In 1964, under the sponsorship of the Air Force Office of Scientific Research (AFOSR), Eugene Garfield discussed the power of using citation data to explore the history of science (Garfield *et al.* 1964). Garfield's notion of devising algorithms to reconstruct the intellectual history of a research field became realized with the introduction of *HistCite*™, a program for mapping connections between cited references from a set of citing publications (Garfield *et al.* 2003). Marx and Bornmann (2013) proposed plotting the references in documents along the time axis as a spectrogram. Variations of such "Reference Publication Year Spectroscopy" (RPYS) can provide, in our opinion, a set of algorithms and visualizations to complement the approach taken by *HistCite*™ for identifying seminal papers, citations classics, the temporal dynamics of historical influences, and other signals in cited reference data (Bornmann *et al.*, 2014; Comins and Hussey 2015b; Comins and Leydesdorff 2016; Leydesdorff *et al.*,2014).

In this paper, we strive to make RPYS techniques accessible to a broader audience of bibliometricians and other researchers by providing an online tool for performing two RPYS analyses interactively: (1) standard RPYS and (2) Multi-RPYS. Standard RPYS can be used to identify the seminal historical works of a research field, whereas Multi-RPYS allows users to visualize the contribution of cited references over the publication history of the citing articles. Specifically, the latter technique enables us to differentiate transiently cited references at a research front from those receiving citations over a sustained period of time as well as to capture the rise and decline of references in citing sets such as in the case of moving research fronts (Comins and Leydesdorff 2016).

Currently, two tools are already available for computing standard RPYS: (1) RPYS.exe (Leydesdorff *et al.* 2014) and (2) CRExplorer.exe (Thor *et al.* 2016). These are downloadable and executable programs for generating standard RPYS outputs. The outputs include a plot of the sum of the cited references by the reference publication year. Additionally, users can create a plot that subtracts such sums for a given reference publication year from the 5-year median to "de-trend" the results, which escalate over time. CRExplorer adds tools for the disambiguation of the cited references, particularly important for evaluation studies.

The tool being introduced here, RPYS i/o, differs from RPYS.exe and CRExplorer.exe in three key ways: (1) it exists online so that users can access interactively; (2) it runs both standard and Multi-RPYS analyses; and (3) it provides direct links to the papers cited via the digital objet identifier (DOI) or a search in Google Scholar. The major limitation to RPYS i/o is that, in the current version, a limit of 15 Mb is set in the size of files that can be submitted for analysis. In our opinion, the three tools can play different roles in the development of bibliometrics. The platform for conducting Multi-RPYS analyses is key for further developments in citation theory: citations have different functions as currency at the research front or as concept-symbols in longer-term developments (Baumgartner and Leydesdorff, 2014; Price, 1970; Small, 2008). A unique feature of RPYS i/o, furthermore, is the direct linking to the referenced documents if a DOI is available or to the search in Google Scholar otherwise.

## 2. Materials and Method

RPYS i/o serves as an online platform for performing standard RPYS and Multi-RPYS analyses. Users utilize one of two clearly marked upload forms to submit full records downloaded in the "Plain Text" format from the Web of Science (WoS) Core Collection for analysis. WoS files can be uploaded to the server (at comins.leydesdorff.net), read and analyzed. (For privacy reasons, files are thereafter disposed.) Our procedures for structuring and analyzing the data utilize the PHP server-side scripting language; visualizations apply the HighCharts JavaScript visualization library (http://www.highcharts.com/, free for non-commercial use and use by non-profit organizations); and searchable HTML tables leverage the Editablegrid JavaScript library (http://www.editablegrid.net/, available under the MIT license). We note that the size limit of 15 MB can be increased if there is demand from the user community. For optimal performance, we recommend using the Chrome or Safari browsers; certain browsers (e.g., Firefox)[3] will not always properly run the web-application.

Prior work discusses in detail the procedures for performing standard RPYS (Comins and Hussey 2015a; Elango *et al.* 2016; Leydesdorff *et al.* 2014; Marx and Bornmann 2013; Marx *et al.* 2014; Thor *et al.* 2016; Wray and Bornmann 2014) and Multi-RPYS analyses (Comins and Hussey 2015b; Comins and Leydesdorff 2016). Briefly, standard RPYS begins by aggregating the cited references from a set of citing publications. The aggregated references are then plotted by the reference publication year. Finally, by taking the difference between the number of cited references for any given reference publication year (e.g., $n$) from the 5-year median of reference publications ($n-2$, $n-1$, $n$, $n+1$, $n+2$). Multi-RPYS extends this procedure by segmenting the set of citing articles by their publication years and performing a standard RPYS analysis for each year under study. The results are then rank-transformed and organized in a heatmap to visualize the dynamic influences of cited references on the citing set. The current version includes referenced publication years from 1900 to1999, but future iterations of the software will expand the choices about this coverage.

To illustrate usage of RPYS i/o, we accessed and downloaded data using the Thomson

---

[3] Firefox, for example, may prompt for continuation of the script after a period of time-out.

Reuters Web of Science Core Collection on January 4, 2016. Our dataset covers 4,024 documents published in the journal *Philosophy of Science* (*PoS*; see Wray and Bornmann, 2015). In total, this dataset includes 36,945 cited references; the file size measures 6.6MB. However, one can use any valid search string in WoS. On a results page from a query, the user should select *Full Record and Cited References,* and export using the option *Save to Other File Formats> Other Reference Software> Plain Text.* The download can then be uploaded to RPYS i/o.

## 3. Results

### 3.1. Standard RPYS

We begin our overview of RPYS i/o by performing a standard RPYS analysis of *PoS* documents using the corresponding upload form and submit button on RPYS i/o (Figure 1). Our choice of analyzing articles from *PoS* is largely due to the fact that prior scholarly research has already utilized RPYS on this source (Wray and Bornmann 2014), so that we can focus on the technical features of RPYS i/o in an already analyzed historical context. The test-set in the required input format is available for download at http://comins.leydesdorff.net/pos.txt .

Once the user submits the data file, it is uploaded to the server, analyzed, and visualized. For a file size of 6.6MB, this entire process should take less than a minute. We note that the resulting graph and data table load in a piecemeal manner, so users can expect to see a data table being populated prior to the rendering of the graphs.

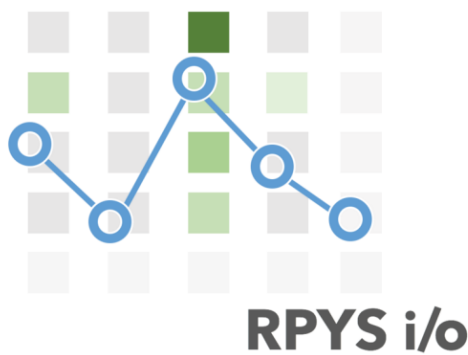

**Figure 1.** Homepage of RPYS i/o, an online platform for running standard RPYS and Multi-RPYS analyses. Users can upload full records up to 15MB exported from the Web of Science Core Collection. The top upload form submits the data to standard RPYS analysis while the bottom form submits the data to Multi-RPYS analysis.

Running the option "Standard RPYS", plots with two sets of values and one data table is generated. The plot contains a grey column chart of the raw frequency count of referenced publication years cited within the set of citing article. The values correspond to numbers shown on the left y-axis. The red spline represents the difference between the raw frequency count of references for any given reference publication year from the 5-year median of reference publication years.

For example, our data set contains 1,339 references published in 1980. When we subtract the median number of cited references from the years 1978, 1979, 1980, 1981 and 1982 from the 1,339 cited references in 1980, this yields a value of 314. Compared to neighboring reference years, this high value reveals 1980 as an important peak year in the intellectual history of documents published in *PoS*. When using the mouse to hover over the graph, a tooltip reveals the raw frequency and difference from median values for any given year (Figure 2). Additionally, users can click-and-drag the mouse to zoom in on regions of the graph. This automatically renders a "reset zoom" button in the top right hand corner of the graph that, when clicked, returns to the original zoom setting. Finally, users can export the visualization by clicking on the menu in the upper right-hand corner of the plot (three grey bars stacked horizontally).

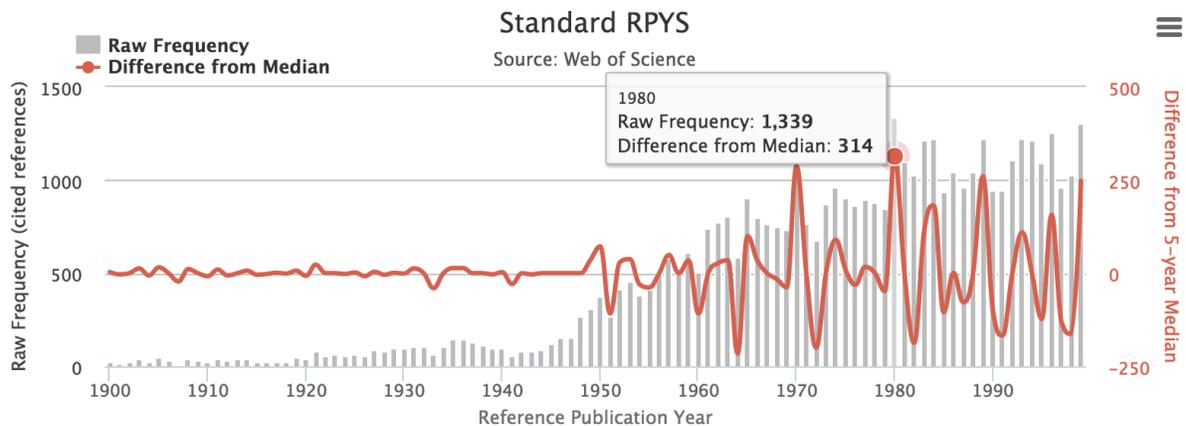

**Figure 2.** Standard RPYS plot generated by RPYS i/o. The left y-axis reveals the raw frequency of cited references (grey columns), the right y-axis shows the difference of cited references of a given year from the 5-year median (red spline) and the x-axis is the reference publication year (1900-1999).

Along with the graph, RPYS i/o generates a searchable data table to help users explore what references are driving the peak behaviors observed in the RPYS plot. Search fields include the Author(s) noted from the WoS cited references section, the reference publication year (RPY), the source of the cited reference, the number of times the document was referenced, and a link if the DOI is available or to a Google Scholar search of the cited reference otherwise.

For example, based on the peak seen in 1980 one can search sort the table for all references cited from 1980 by typing *RPY1980* into the search box. (Users do not need to hit enter, since searches are performed iteratively per key stroke.) Furthermore, by clicking on the table headers, one can sort the table results in ascending or descending order; in this case, sorting by times referenced reveals the most common references are to Van Fraassen's (1980) book *The Scientific Image*. We can further refine the search by adding the letters *fra* to our original search term of *RPY1980* (Figure 3). Doing so, we see several variations of references to Van Fraassen (1980) which was cited (141 + 11 + 10 + 3 * 1 =) 165 times in *PoS*. For further disambiguation of the cited references, the user is referred to CRexplorer (at http://crexplorer.net; Thor *et al*., 2016; cf. Leydesdorff, 2008: 285, Table 4).

On the right hand side of the table, links will either directly send the user to the articles if a DOI is available in this set of references or, in the absence of a DOI, it will search Google Scholar for the cited reference. One current limitation of RPYS i/o at present is that, to maintain reasonable performance, table rows shown are limited to 40; however, the search feature provides ample occasion for exploring key reference publication years and cited references for a citing set.

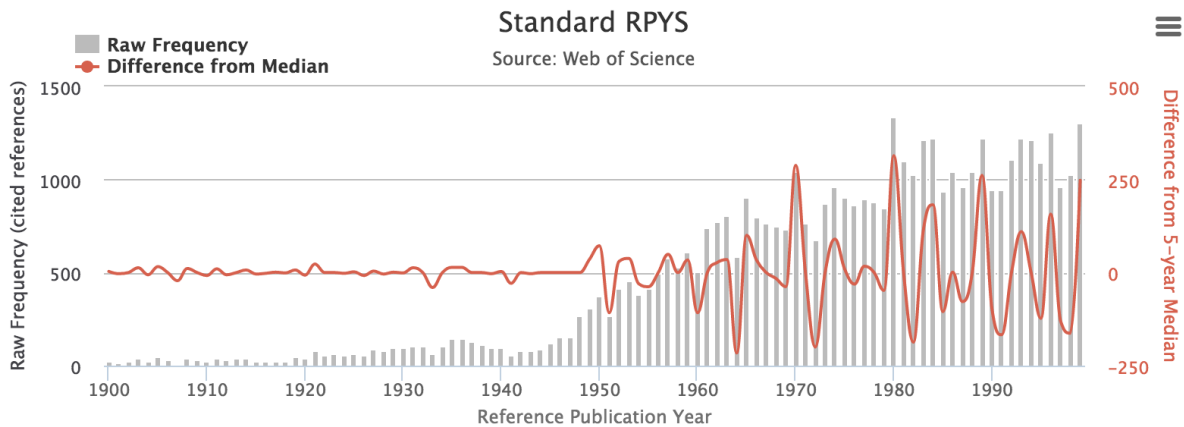

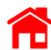

**Figure 3.** Utilizing the search table to explore standard RPYS results in RPYS i/o. In this example, we search for RPY1980 to refine the table to references cited from 1980 and add the letters *fra* to narrow in on references to Van Fraassen's *The Scientific Image*, the most cited reference from 1980 for documents published in *Philosophy of Science*.

### 3.2. Multi-RPYS

Let us now demonstrate the usage of RPYS i/o to run a Multi-RPYS analysis. (Users can return to the homepage by clicking either the *Home* link in the top left-hand corner or the red icon of a house beneath the plot.) In this case, the upload form at the bottom of http://comins.leydesdorff.net/ should be used. Again, the user's data will be uploaded, analyzed, and then deleted from the server. In essence, the Multi-RPYS plot runs a distinct standard RPYS analysis for the set of citing documents segmented by their publication years. The resulting heatmap organizes the data by the publication year of the citing publications on the y-axis, publication year of the cited references on the x-axis and the color of the cells in the heatmap are determined by the rank-transformed value of the difference from the 5-year median (Figure 4). Visualizing the data in this way allows users to identify references that made enduring contributions to the citing documents (e.g., often seen as bands or "pillars" in the heatmap) as well as the potential to find references whose importance to the citing set is rising or waning (Comins and Leydesdorff 2016).

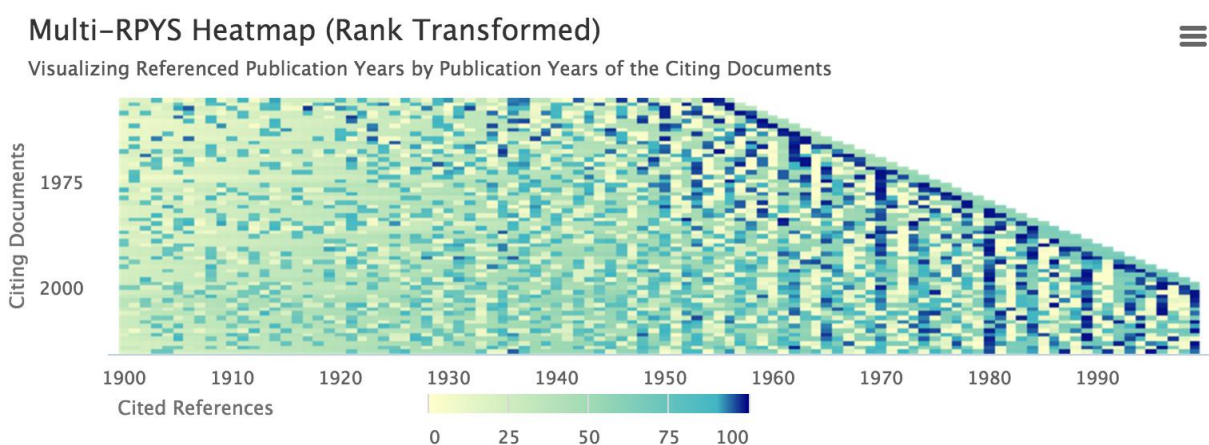

**Figure 4.** Multi-RPYS visualization of documents published in *Philosophy of Science*. Bands in the heatmap correspond to publications with enduring contributions to a field, such as Kuhn's *The Structure of Scientific Revolutions* in 1970 and Van Fraassen's *The Scientific Image* in 1980.

Again, the functionality of the table search provides an excellent way to explore the intellectual history for this citing set. In addition to the searchable fields seen for the standard RPYS analysis users can search for the publication year of the citing set. The band seen in 1980 suggests the enduring importance of references published in this year. Using the table, we can identify that this is driven by Van Fraassen's (1980) book entitled *The Scientific Image*. We follow up on the example from earlier by assessing how Van Fraassen's work continues to be cited by recent documents within the citing set. For example, by searching for *RPY1980 fra*

*CPY201* we find that *The Scientific Image* is referenced 24 times by documents published in 2010, 2011, 2012, 2013, 2014 and 2015 (Figure 5). One can compare the figures for these years with the much larger numbers including all years of citation in the legend to Figure 3.

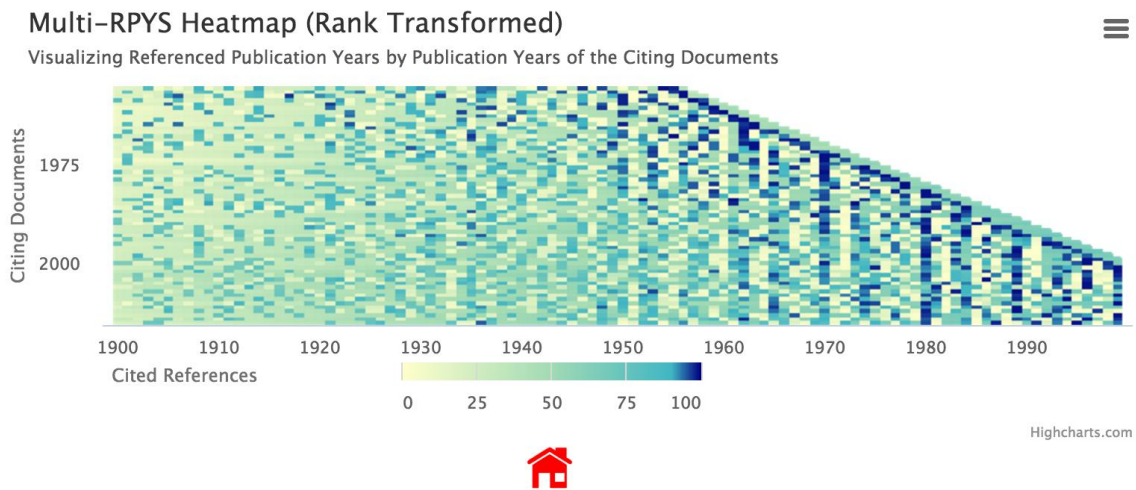

**Figure 5.** Using the table search function, users can refine and explore the interactions between reference publication years and publication years of the citing documents. In the example here, we highlight references to Van Fraassen's work from 1980 cited since 2010 by search *RPY1980 fra CPY201*.

One can compare Van Fraassen's citation rates in this domain, with those of Thomas Kuhn's (1970) *The Structure of Scientific Revolutions,* which contributes with 98 citations to the darkening of the 1970 bar until the early 1980s (citing). Referencing this book fades away to only six times during the period 2010-2015. However, one can add four references to the first edition of the same book albeit without the important *Preface* to the second edition, dated 1969.

The 1962-book is referenced 68 times in *PoS* so that Kuhn's book is cited 164 times in total and thus of the same order as Van Fraassen (1980) that happens to be cited 165 times. The more recent citation rates of the two books, however, are significantly different. On the one hand, this may illustrate "obliteration by incorporation", since Kuhn's revolutionary notions of paradigms and evolutionary cycles have been so widely accepted that one no longer needs to provide the reference (Cozzens, 1989; Garfield, 1975). On the other hand, it may indicate a turn away from the larger historical context provided by Kuhn as a historian of science, to the more internalist orientation of Van Fraassen (1980)'s book (Frodeman and Briggle, 2016).

A final point that we wish to note in Figure 5 is the dark line along the diagonal in the upper-right part of the drawing. This line indicates the research front of short-time citation, that is, in the first two or three years before publication. One does not expect a lot of this current referencing in *PoS* (given the nature of this field), but obviously authors feel themselves obliged to provide references to the current environment; for example, in order to indicate the social and scientific relevance of the new knowledge claim. Baumgartner and Leydesdorff (2014) noted that most indicator studies focus on these recent citations, which may be absent or insignificant in the development of scholarly traditions other than the laboratory (Price, 1972).

## 4. Discussion

In this brief communication, we introduced a new online platform, RPYS i/o, for conducting interactively both standard and Multi-RPYS analyses. The contribution to the information and library sciences is mainly methodological; but using examples from the philosophy of science, we could show how interesting questions can be further developed by

Page 12 of 15

using this tool. For example, the issue of the influence of the two editions of Kuhn's (1962 and 1970) book came saliently to the fore. Multi-RPYS enables us to address the distinction between citations at the research front and citations codified into concept symbols (Small, 1978), at the risk of being "obliterated by incorporation" (Garfield, 1975). In our opinion, these different functions of citation merit further attention, because of their important consequences for both theorizing about citation and the development of indicators in evaluation studies (Leydesdorff and Milojević, 2015).

The tool provides flexible options for analyzing and searching cited reference results, visualizing standard RPYS and Multi-RPYS outputs, and exporting graphics for later usage. Such tools aid scholars in studying the intellectual history of fields by identifying seminal papers and notable cases where a paper's influence has persisted, grown, or waned over time. We hope that, in concert with tools such as CRExplorer.exe and RPYS.exe, more bibliometricians will contribute to maturing algorithms for reconstructing the intellectual history of a field. In this context, we also wish to mention CitNetExplorer (at http://www.citnetexplorer.nl/) that is developed by researchers at the Centre for Science and Technology Studies (CWTS) in Leiden as a much needed follow-up in the *HistCite™* tradition of analyzing citation networks over time (Van Eck and Waltman, 2014; cf. Leydesdorff *et al.*, 2016, forthcoming).

## 5. Acknowledgements

We thank Lutz Bornmann and George Chacko for feedback on an earlier version of this manuscript.Page 13 of 15

Leydesdorff, L., and Milojević, S. (2015). The Citation Impact of German Sociology Journals: Some Problems with the Use of Scientometric Indicators in Journal and Research Evaluations. *Soz. Welt*. 66, 193-204.

Marx, W., and Bornmann, L. (2013). Tracing the origin of a scientific legend by reference publication year spectroscopy (RPYS): the legend of the Darwin finches. *Scientometrics*. 99, 839–844.

Marx, W., Bornmann, L., Barth, A., and Leydesdorff, L. (2014). Detecting the historical roots of research fields by reference publication year spectroscopy (RPYS). *J. Assoc. Inf. Sci. Tech*. 65, 751–764.

Price, D. de Solla (1970). Citation Measures of Hard Science, Soft Science, Technology, and Nonscience. In C. E. Nelson and D. K. Pollock (Eds.), *Communication among Scientists and Engineers* (pp. 3-22). Lexington, MA: Heath.

Small, H. (1978). Cited documents as concept symbols. *Soc. Stud. Sci*. 8, 113-122.

Thor, A., Marx, W., Leydesdorff, L., and Bornmann, L. (2016). Introducing CitedReferencesExplorer (CRExplorer): A program for Reference Publication Year Spectroscopy with Cited References Disambiguation. *ArXiv*. 1601.01199.

van Eck, N. J., and Waltman, L. (2014). CitNetExplorer: A new software tool for analyzing and visualizing citation networks. *J. Informetr*. 8, 802-823.

Van Fraassen, B. C. (1980). *The scientific image*. Oxford/New York: Oxford University Press.

Wray, K. B., and Bornmann, L. (2015). Philosophy of science viewed through the lens of "Referenced Publication Years Spectroscopy"(RPYS). *Scientometrics*. 102, 1987-1996.